\documentclass[pdflatex,sn-mathphys-num]{sn-jnl}
\DeclareUnicodeCharacter{2223}{\mid}

\usepackage{graphicx}%
\usepackage{multirow}%
\usepackage{amsmath,amssymb,amsfonts}%
\usepackage{amsthm}%
\usepackage{mathrsfs}%
\usepackage[title]{appendix}%
\usepackage{xcolor}%
\usepackage{textcomp}%
\usepackage{manyfoot}%
\usepackage{booktabs}%
\usepackage{algorithm}%
\usepackage{algorithmicx}%
\usepackage{algpseudocode}%
\usepackage{listings}%

\theoremstyle{thmstyleone}%
\theoremstyle{thmstyletwo}%

\theoremstyle{thmstylethree}%

\begin{document}

\title{Deep Generative Learning of Magnetic Frustration in Artificial Spin Ice from Magnetic Force Microscopy Images}


\author[1,2]{\fnm{Arnab} \sur{Neogi}}\email{aneogi2@lanl.gov}
\equalcont{*These authors contributed equally to this work.}
\author[2]{\fnm{Suryakant} \sur{Mishra}}\email{mishras@lanl.gov}

\equalcont{*These authors contributed equally to this work.}

\author[3]{\fnm{Prasad P} \sur{Iyer}}
\author[3]{\fnm {Tzu-Ming} \sur{Lu}}
\author[3]{\fnm {Ezra} \sur{Bussmann} }
\author[1,2]{\fnm{Sergei} \sur{Tretiak}}
\author*[2]{\fnm{Andrew Crandall} \sur{Jones}}\email{acj@lanl.gov}

\author*[1,2]{\fnm{Jian-Xin} \sur{Zhu}}\email{jxzhu@lanl.gov}

\affil[1]{\orgdiv{Theoretical Division, Los Alamos National Laboratory, Los Alamos, New Mexico 87545, United States}}

\affil[2]{\orgdiv{Center for Integrated Nanotechnologies, Los Alamos National Laboratory, Los Alamos, New Mexico 87545, United States}}

\affil[3]{\orgdiv{Center for Integrated Nanotechnologies, Sandia National Laboratory, Albuquerque, New Mexico 87185, United States}}


\abstract{Increasingly large datasets of microscopic images with atomic resolution facilitate the development of machine learning methods to identify and analyze subtle physical phenomena embedded within the images. In this work, microscopic images of honeycomb lattice spin-ice samples serve as datasets from which we automate the calculation of net magnetic moments and directional orientations of spin-ice configurations. In the first stage of our workflow, machine learning models are trained to accurately predict magnetic moments and directions within spin-ice structures. Variational Autoencoders (VAEs), an emergent unsupervised deep learning technique, are employed to generate high-quality synthetic magnetic force microscopy (MFM) images and extract latent feature representations, thereby reducing experimental and segmentation errors. The second stage of proposed methodology enables precise identification and prediction of frustrated vertices and nanomagnetic segments, effectively correlating structural and functional aspects of microscopic images. This facilitates the design of optimized spin-ice configurations with controlled frustration patterns, enabling potential on-demand synthesis.
}




\maketitle

\section{Introduction}\label{sec1}


Artificial spin ice (ASI) systems are engineered arrays of nanoscale magnetic elements designed to mimic the frustrated spin configurations of natural spin-ice materials \cite{skjaervo2020advances}. These patterned nanomagnet lattices provide a versatile platform for exploring fundamental magnetic phenomena that were previously observed only in bulk spin systems or theoretical models \cite{gilbert2014emergent},\cite{lendinez2019magnetization},\cite{zhang2013crystallites}. In particular, ASI exhibits signatures of magnetic frustration, characterized by a large degeneracy of low-energy configurations and the emergence of excitations resembling magnetic monopoles, all governed by tunable geometric constraints \cite{gilbert2014emergent},\cite{lendinez2019magnetization}. Studying these effects in ASI has broadened our understanding of frustration and emergent magnetism under controllable settings, bridging the gap between theoretical models and real materials.
Beyond fundamental physics, ASI has promising applications as  a platform for new magnetic and spintronic technologies. By tailoring the geometry and coupling of its nanomagnets, an ASI array can function as a reprogrammable magnonic crystal with tunable spin-wave spectra for microwave signal processing and quantum information devices \cite{skjaervo2020advances},\cite{chumak2022advances},\cite{krawczyk2014review}. ASI-based networks have also been proposed for unconventional computing architectures. For example, specific magnetically frustrated nanomagnet arrangements can be used for logic gate implementations or neuromorphic computing
\cite{sievers2012quantitative},\cite{hu2023distinguishing}. Moreover, the ability to manipulate ASI’s emergent magnetic monopole defects using external magnetic fields, electric currents, or thermal gradients \cite{castelnovo2008magnetic},\cite{morris2009dirac},\cite{nisoli2010effective} paves the way for in-situ control of magnetic charge transport, enabling novel active metamaterials and next-generation functional devices.

\begin{figure}[h!]
\centering
\includegraphics[width=1.0\textwidth]{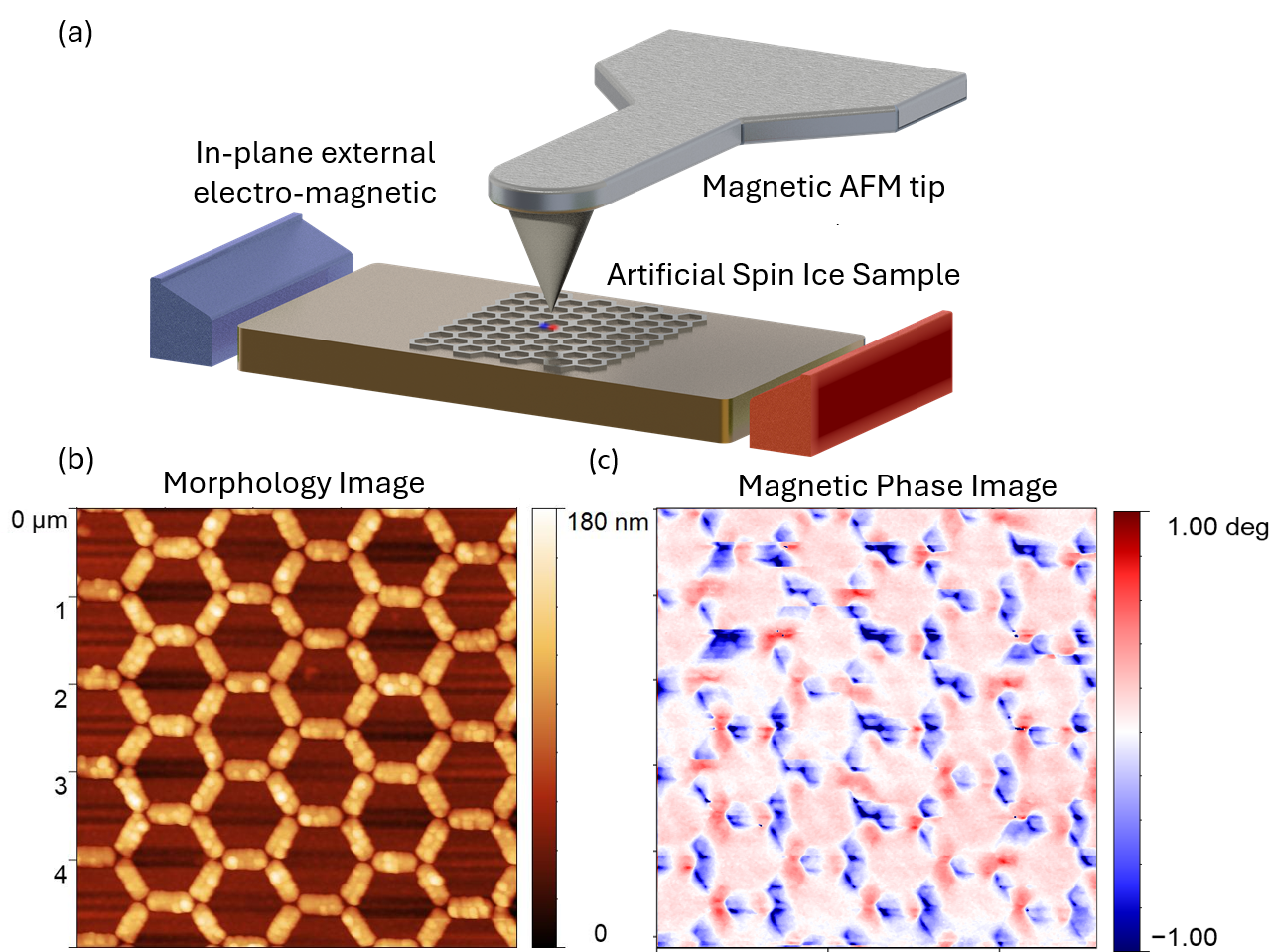}
\caption{(a) A schematic representation of magnetic-AFM illustrating the experimental setup for ASI sample measurement, along with its corresponding (b) morphology and (c) magnetic phase image. The blue and red colors in the phase image represent the North and South poles of the magnetic bars, respectively.}\label{experimental_workflow}
\end{figure}

Magnetic Force Microscopy (MFM) is a powerful technique for exploring and characterizing complex magnetization patterns in ASI systems. MFM is a variant of atomic force microscopy that uses a magnetized probe tip to scan above the sample surface and sense the stray magnetic fields produced by individual nanomagnets. This method provides high-resolution, non-destructive imaging of local magnetic configurations in an ASI lattice \cite{ladak2010direct}. Unlike bulk magnetometry methods that measure only a sample’s aggregate response, MFM directly visualizes the magnetization of individual bars and the state of each vertex, enabling real-space analysis of spin interactions at the nanoscale. The technique’s strong sensitivity to nanoscale stray fields makes it particularly effective for capturing magnetization reversal processes, thermally activated excitations, and field-induced phase transitions in ASI systems \cite{gartside2018realization}.
 Figure \ref{experimental_workflow} illustrates an MFM measurement on an ASI sample. In Fig. \ref{experimental_workflow}a, a schematic of the magnetic-AFM setup shows a magnetized tip scanning over the patterned nanomagnet array to detect local magnetic forces. Figures \ref{experimental_workflow}b and \ref{experimental_workflow}c exemplify the two complementary outputs of the measurement: the sample’s surface morphology (topographic image) and the corresponding magnetic phase image, respectively. In the MFM phase image, each nanobar’s stray field produces a bipolar contrast with blue and red regions indicating the north and south magnetic poles at opposite ends of the bar. This contrast explicitly reveals the magnetization direction of every element. This direct imaging of nanoscale dipole orientations and polarities enables immediate identification of the magnetic state of each nanomagnet in the ASI lattice.
Continued improvements in MFM instrumentation have further enhanced its ability to probe ASI. Modern MFM probes with higher spatial resolution and lower magnetic moment coatings minimize tip-induced perturbations while increasing image contrast, now resolving magnetic features as small as 10 nm \cite{kazakova2019frontiers}. Additionally, the integration of variable in-plane magnetic field sample holders into MFM systems enables real-time observation of ASI magnetization dynamics, capturing field-driven transitions in situ during imaging. These advances have accelerated ASI research by allowing in-depth exploration of various topological or frustration-induced configurations and by opening the door to spin‐logic experiments under realistic conditions \cite{sievers2012quantitative}. As MFM continues to evolve, it remains a critical tool for understanding and harnessing the unique properties of ASI systems for future technological applications.
While MFM provides rich visual data on ASI, analyzing the resulting images to quantify spin configurations can be challenging and labor-intensive. The increasing volume of high-resolution microscopy data has motivated the use of machine learning (ML) techniques in condensed matter physics \cite{kalinin2015big}\cite{ziatdinov2017deep} to automate image analysis and feature extraction. In the context of ASI, however, traditional image-processing methods often rely on heuristic thresholds or manual identification of nanomagnets \cite{wang2006artificial}, which can introduce errors due to noise, instrumental artifacts, and user bias. This limitation creates a need for robust data-driven approaches capable of reliably extracting meaningful physical information, such as each nanomagnet’s moment orientation and the energy or “frustration” state of each vertex—from MFM images. Such automated analysis is particularly important given the highly complex, correlated magnetization patterns that ASI exhibits \cite{nisoli2013colloquium}, which can be difficult to interpret with conventional techniques.

In this work, we address these challenges by integrating MFM imaging with an unsupervised deep learning approach for quantitative analysis of spin-ice configurations. We employ a variational autoencoder (VAE) to learn latent feature representations from MFM phase images of a honeycomb ASI lattice. This data-driven approach enables automated identification of each nanomagnet’s magnetic moment direction and the detection of frustrated (high-energy) vertex states directly from the images. The VAE model effectively captures the high-dimensional correlations in the ASI’s magnetic phase patterns while reducing experimental noise and segmentation errors. Furthermore, the learned latent space can be used to generate synthetic MFM images that replicate experimental features, offering deeper insight into how frustration manifests and can be controlled in these networks. By capturing subtle signatures of emergent frustration that might be missed by manual analysis, our framework provides a powerful tool for discovering and designing optimized spin-ice states with tailored properties, addressing the highly tunable interactions and reprogrammable magnetism of ASI systems \cite{nisoli2007ground}. Ultimately, by combining machine learning with advanced microscopy, this approach provides a scalable and precise platform for characterizing ASI systems, leading to on-demand “frustration engineering” in magnetic metamaterials \cite{kempinger2021field} and advancing spin-based information processing technologies.

\begin{figure}[h!]
\centering
\includegraphics[width=1.0\textwidth]{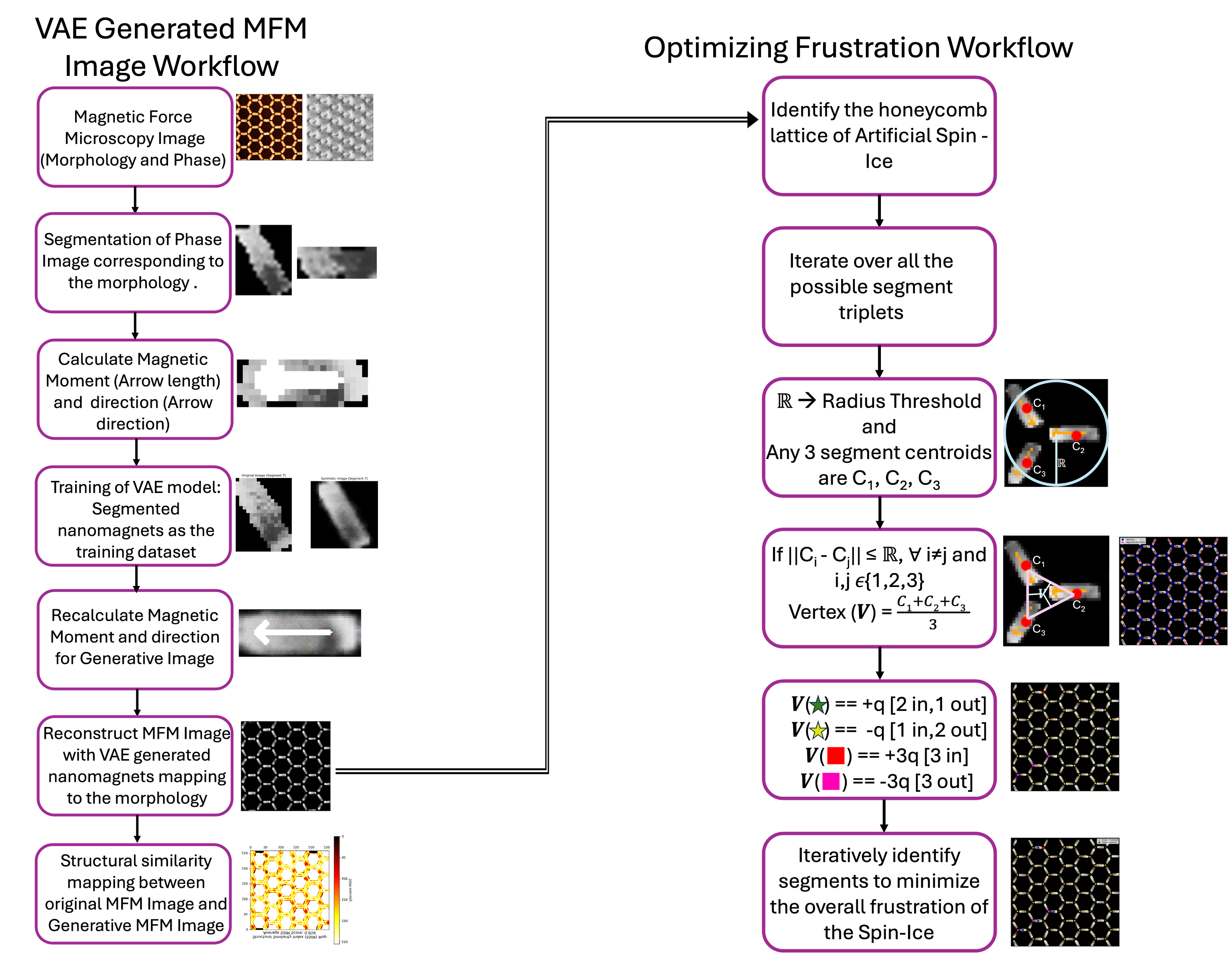}
\caption{The scheme of the overall workflow comprising two distinct components. First, we apply a segmentation algorithm to partition the MFM image into distinct magnetic domains. From the original MFM segments, we train a VAE model to generate synthetic MFM segments. In this representation, the direction of each arrow indicates the nano-magnet dipole direction, while the arrow length corresponds to the normalized magnetic moment of the dipole. The segment position and orientation are retrieved from the original MFM image and reconstructed. The second stage of our workflow focuses on classifying the vertices of a honeycomb lattice spin-ice sample into high energy/highly frustrated ($\pm 3q$) and low energy/less frustrated ($\pm q$) vertices. Furthermore, our algorithm  not only identifies the high energy monopoles of the spin-ice lattice but also determines which segments, if toggled, would minimize the overall rustration.}\label{ML_workflow}
\end{figure}


\section{Methods}\label{method}

\subsection{Magnetic Force Microscopy Measurements}\label{method_exp}

Nanoscale topography and MFM shift images were collected using an Atomic Force Microscope system (DriveAFM, Nanosurf AG). Measurements were performed using \lq lift mode' to record phase contrast variations associated with the magnetic phase of the ASI sample surface.  First, a measurement of the topography of the sample surface was performed using tapping-mode feedback of the atomic force microscopy probe via a line trace and retrace scan.  Next, to reduce the effect of magnetic phase artifacts associated with the sample topography, the probe is then separated by a controlled distance from the surface (70 nm), and a subsequent, elevated, trace, and retrace line scan is performed while recording the phase shift of the AFM probe. The external, in-plane magnetic field was produced using a variable magnetic field sample holder (VMFSH, NanoSurf) with a base that contains the permanent magnets and an integrated calibrated Hall sensor.  Magnetic atomic force microscopy probes were used (Co/Cr coated Bruker MESP-LM-V2) with a spring constant of 2.2 N/m, and resonance frequency of $\sim$65 kHz.  These probes were magnetized in an out-of-plane direction using a permanent magnet prior to measurements.


\subsection{Automating Net Magnetization Calculation} \label{method_magnetization_arrows}

To automate the calculation and analysis of net magnetization in ASI systems, we initiate an image processing pipeline that extracts and uniquely labels individual nanomagnetic segments from the MFM morphology image. The accurate segmentation forms the foundation for computing net magnetic moments in subsequent stages of the workflow.

\begin{figure}[h!]
\centering
\includegraphics[width=1.0\textwidth]{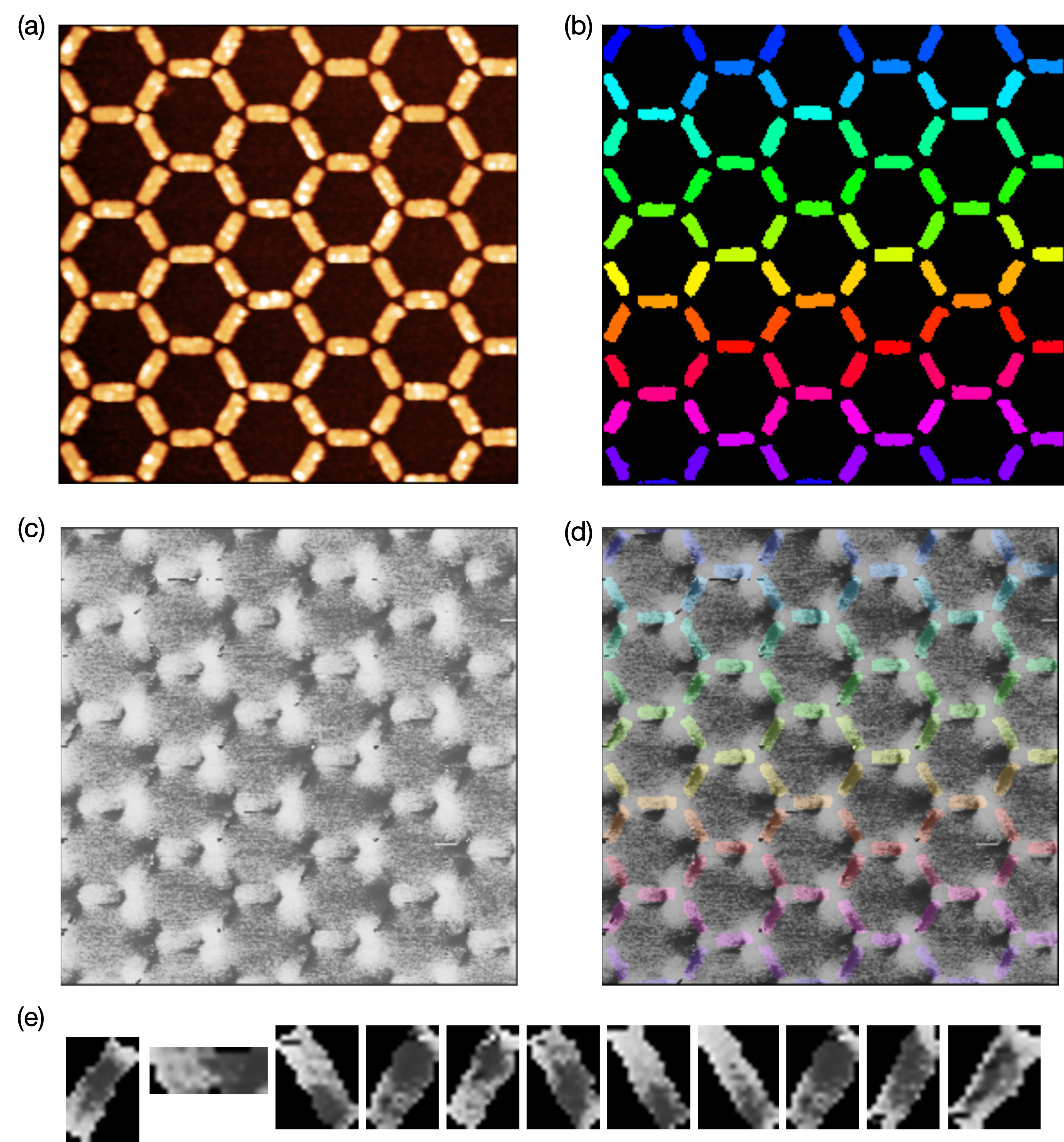}
\caption{(a) Morphology of the ASI system. (b) Uniquely colored segmentation of  the morphology capturing the shape and position of each segments. (c) Gray-scale MFM  phase image. (d) Uniquely colored morphological segmentation image overlayed on the gray-scale MFM image. (e) Examples of segments from the MFM image segmented based on the overlay morphological image representing nano-magnets 
 contributing to the overall layout of the artificial spin ice sample.}\label{morphology_&_bit_segmentation}
\end{figure}

\subsubsection{Segmentation of Nanomagnets}\label{segmentation method}

The input morphology image (Fig. \ref{morphology_&_bit_segmentation}(a)) is first converted to grayscale, followed by the application of a binary thresholding mask to distinguish magnetic regions from the background. This step ensures that only relevant segments are considered. Euclidean distance transform refines the segmentation by enhancing the separation between foreground structures and suppressing noise, thereby improving segmentation accuracy. It converts the binary mask into a grayscale map, where brighter intensities represent regions farther from the boundary keeping only high-intensity regions, ensuring robust object detection.

To identify individual nanomagnetic segments, Connected Components Analysis (CCA) from the OpenCV2 computer vision library \cite{opencv_library} is employed. This step assigns unique labels to distinct magnetic structures. However, in cases where segments are closely packed, CCA alone may not effectively separate overlapping structures. To address this, the Watershed Segmentation Algorithm \cite{beucher1979use} is applied to refine boundaries further. The watershed algorithm uses gradient information to distinguish touching objects by identifying high-gradient regions as boundaries, ensuring each nanomagnet is distinctly labeled and effectively separated.

Following segmentation, contours are extracted for each labeled segment using OpenCV’s "findContours()" function which identifies the outer boundary of each detected region by tracing along intensity changes in the binary mask, where each segmented nanomagnet has a unique label. OpenCV’s "cv2.findContours()" is applied to this mask with external retrieval mode ("cv2.RETR\textunderscore EXTERNAL") to detect only the outermost contour, ignoring any nested inner contours.  
The simplified chain approximation reduces redundant points and store only the essential boundary points of each segment.  

Only segments with an area exceeding a predefined threshold are retained to filter out small, incomplete, or boundary-clipped structures. These selected segments are then assigned unique colors using an HSV colormap, enhancing visual differentiation.  
The center coordinates of each segment are determined using the mean position of the contour points, and each segment is numerically labeled for reference and tracking. Finally, the segmentation map is compared with the original morphology image to validate the accuracy of the detected structures.  

The individual segments are reassembled for full segmentation visualization, which ensures that the extracted segments are correctly placed within the original morphology image. The individual segmented nanomagnets, stored as separate image files, are sorted and iteratively placed back into the original layout. To correctly reposition each segment, the corresponding label position is retrieved from the segmentation mask mapping segments to their original positions.

 The MFM image with magnetic contrast data, is segmented similarly using the  reconstructed segmentation map, such that each nanomagnetic region is extracted with accurate boundaries. This method aligns the spatial regions in the MFM image with the segmented morphology image, allowing further analysis of magnetic properties at the nanoscale and we validate the segmentation process by comparing the extracted segment count with the number of unique labels.
This step ensures that the MFM image is segmented accurately, preserving the spatial and morphological characteristics of each nanomagnet for further magnetic analysis.

\begin{figure}[h!]
\centering
\includegraphics[width=1.0\textwidth]{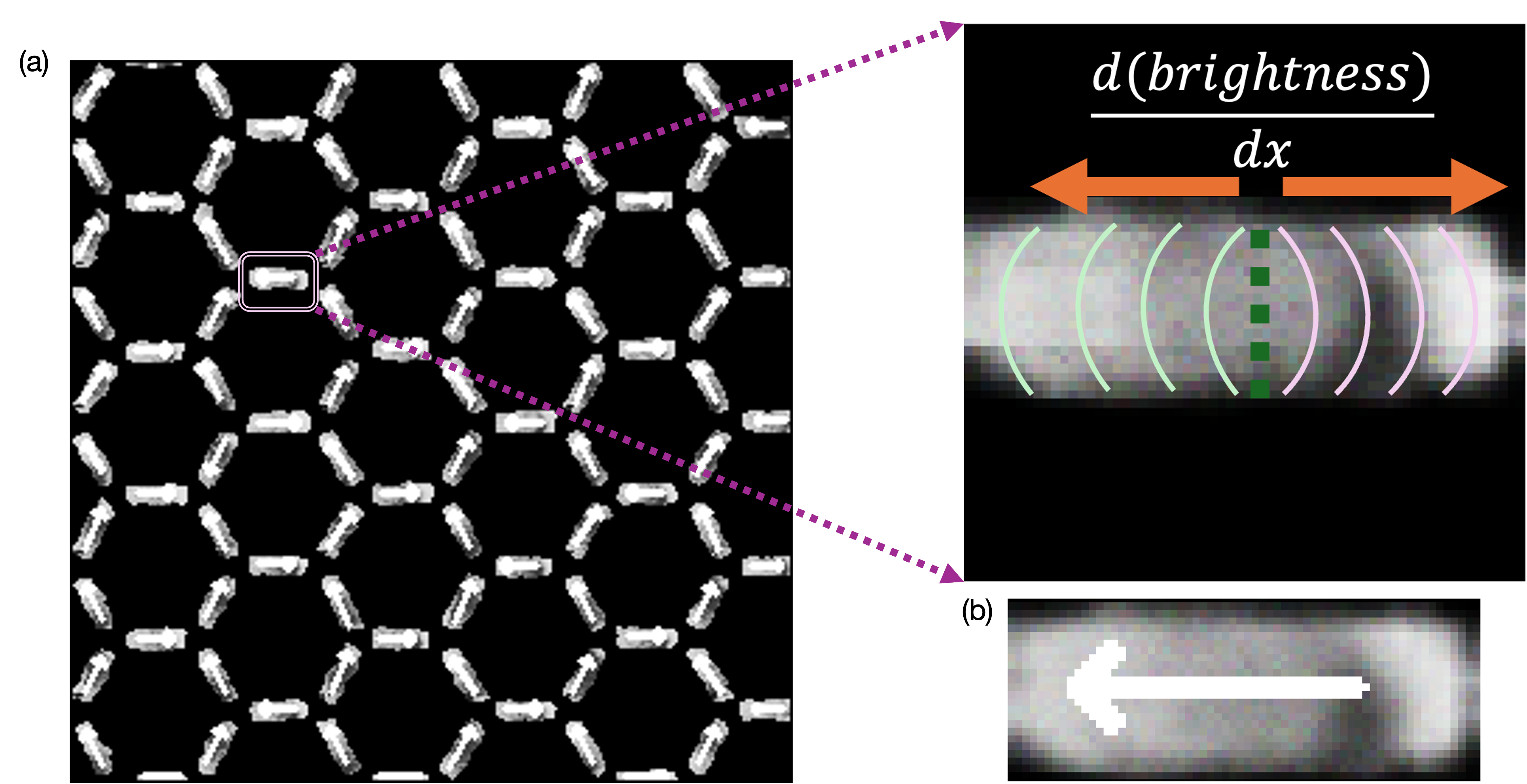}
\caption{(a) The segmented nano-magnets from Figure \ref{morphology_&_bit_segmentation}(e) are reconstructed with position of each segment retrieved from the labeled segment data of the morphology of spin-ice sample. The right side is a zoomed in nano-magnet with artifact, classified as an adversary attack on the expected dipole image. An expected dipole nano-magnet image should have a bright(North) and a dark end (South). The segment is subdivided into two halves, each half is radially scanned calculating brightness gradient. The gradient decreases along the South end of the nano-magnet while increases along the North end. Reaching the artifact radial area, the decreasing brightness gradient experiences sudden increase ($>$ threshold contrast) and classifies the rest of the segment as an adversary. The erroneous part is not included in calculating the magnetic moment calculation. (b) Validating the calculation process, the arrow of the zoomed in segment points towards the North neglecting the adversary.}\label{Calculaion method of magnetic moment}
\end{figure}

\subsubsection{Magnetic Moment Calculation}\label{magnetic moment method}

Next we calculate the magnetic moment direction and strength of each nanomagnet from the segmented MFM image which involves segment extraction, brightness-based dipole determination, and arrow visualization to indicate the moment direction. The segment image is converted to grayscale, and binary thresholding is applied to create a mask that separates the magnetic structure from the background. Thresholding isolates the nanomagnet segment by setting pixel intensities above a defined threshold to white (255) and others to black (0). The contours of the binary mask are extracted to identify the outer boundary of the nanomagnet. The bounding box of the segment is calculated using contour fitting, providing the position (x, y), width, and height and the diagonal length of the bounding box is computed to determine the appropriate arrow size. The orientation of each segment is estimated by fitting an ellipse to the extracted contour. The major axis of the ellipse represents the longest dimension of the nanomagnet, which is used as the magnetization direction.
The orientation angle of the ellipse is extracted and rotated by 90$^{\circ}$ to align with the expected moment direction. 

To establish the direction of the magnetic moment, our algorithm analyzes brightness variations across the nanomagnet. The segment is split into two halves along the major axis of the ellipse and brightness values from both halves are extracted, and sudden intensity changes are identified using gradient analysis. The radial gradient of brightness are calculated starting from the separation line of towards the edge of the contour. If there is a sudden intensity change along the major axis of the ellipse, we identify it as the artifact of the dipole as shown in the zoomed in segment image in Fig. \ref{Calculaion method of magnetic moment}, considering that each segment is expected to have a brightness gradient from bright to dark, inferring bright side as the head of the dipole.  
Filtered brightness values are computed to reduce noise effects and capture a more reliable contrast between the two sides.
The side with higher brightness is inferred to be the head of the dipole, as MFM contrast reflects magnetization variations. 
If $I(x,y)$ be the grayscale intensity at pixel (x,y). The brightness of each half is computed as:

\begin{equation}
    B_{{side}_{1}} = \frac{1}{N_1} \sum_{(x,y)\epsilon side_1} I(x,y)
\end{equation}

\begin{equation}
    B_{{side}_{2}} = \frac{1}{N_2} \sum_{(x,y)\epsilon side_2} I(x,y)
\end{equation}

where:
$B_{{side}_{1}}$ and $B_{{side}_{2}}$ are the mean brightness values of the two halves, $N_1$ and $N_2$ are the number of pixels in each half.

The brightness difference  between the two halves is normalized to scale the arrow length, so that it varies proportionally with the MFM contrast.
The arrow length is constrained within a defined range to maintain visual clarity and the direction of the arrow is determined based on the brighter half, with the arrowhead pointing towards it.

The contrast strength ($C_s$), which defines magnetization strength, is given by:

\begin{equation}
C_s = |B_{{side}_{1}} - B_{{side}_{2}}|
\end{equation}
Since pixel intensities in grayscale images range from 0 to 255, the contrast is normalized: $C{_s}^{norm} = \frac{C_s}{255}$ so that $C{_s}^{norm}$ ranges from [0,1]. With normalizing, the contrast strength is scaled relative to the maximum brightness range. The arrow length is then linearly scaled between the predefined minimum and maximum arrow lengths such that:

\begin{equation}
    L_{arrow} = L_{min} + C{_s}^{norm} . (L_{max} - L_{min})
\end{equation}
The $L_{max}$ and $L_{min}$ are chosen to be 90\% and 55\% respectively of the diagonal length of the identified contour($L_{diag}$) where $L_{diag} = \sqrt{w^2+h^2}$. The final arrow direction  is determined by identifying which half of the segment is brighter:

\begin{equation}
    \theta_{\text{arrow}} =
\begin{cases} 
\theta_{\text{major}} + 90, & \text{if } B_{\text{side}_1} > B_{\text{side}_2} \\
\theta_{\text{major}} + 270, & \text{otherwise}
\end{cases}
\end{equation}

The $\theta_{\text{major}}$ determines the major axis of the fitted ellipse and if $Side_1$ is brighter the arrow points along the major axis, otherwise, the arrow is rotated by 180$^{\circ}$ to indicate reversal.

This magnetic moment and strength is iterated over all the segmented nanomagnet from the MFM image and then reassembled following the morphology image. The net magnetic moment, $M_{\text{net}}$, of the ASI sample is  calculated as the sum of arrow lengths, where the arrow length represents the relative strength of the magnetic moment for each segment.

\begin{equation}
M_{\text{net}} = \sum_{i=1}^{N} L_{\text{arrow}, i}
\end{equation}

here $L_{\text{arrow}}$ is the arrow length corresponding to the magnetic moment strength of the i-th segment and N is the total number of segments.

To generalize the scaling of the net magnetic moment from arbitrary units (a.u.) to physical units(mT) use the unit used in experiment, we define a scaling transformation that maps the computed image-based net magnetization to the experimentally measured range.
From the experimental section, we define the real magnetic moment range: 
  $Real_{Range}= M_{max} - M_{min}$.
From image processing algorithm, we define the arbitrary range:
  $ Arbitrary_{Range} = L_{max} - L_{min}$ 
Thus the scaling factor is computed as:
\begin{equation}
    S = \frac{Real_{Range}}{Arbitrary_{Range}}
\end{equation}

\begin{equation}
M_{\text{real}} = \left( \frac{M_{\max} - M_{\min}}{L_{\text{max}} - L_{\text{min}}} \right) \cdot M_{\text{image}}
\end{equation}

\begin{equation}
M_{\text{real}} = S \cdot M_{\text{image}},
\end{equation}

 where $M_{real}$ is the real net magnetic moment and $M_{image}$ is the net magnetic moment computed from image processing (in arbitrary units, a.u.).

\subsubsection{Variational Auto Encoder Architecture}\label{VAE method}

\begin{figure}[h!]
\centering
\includegraphics[width=1.0\textwidth]{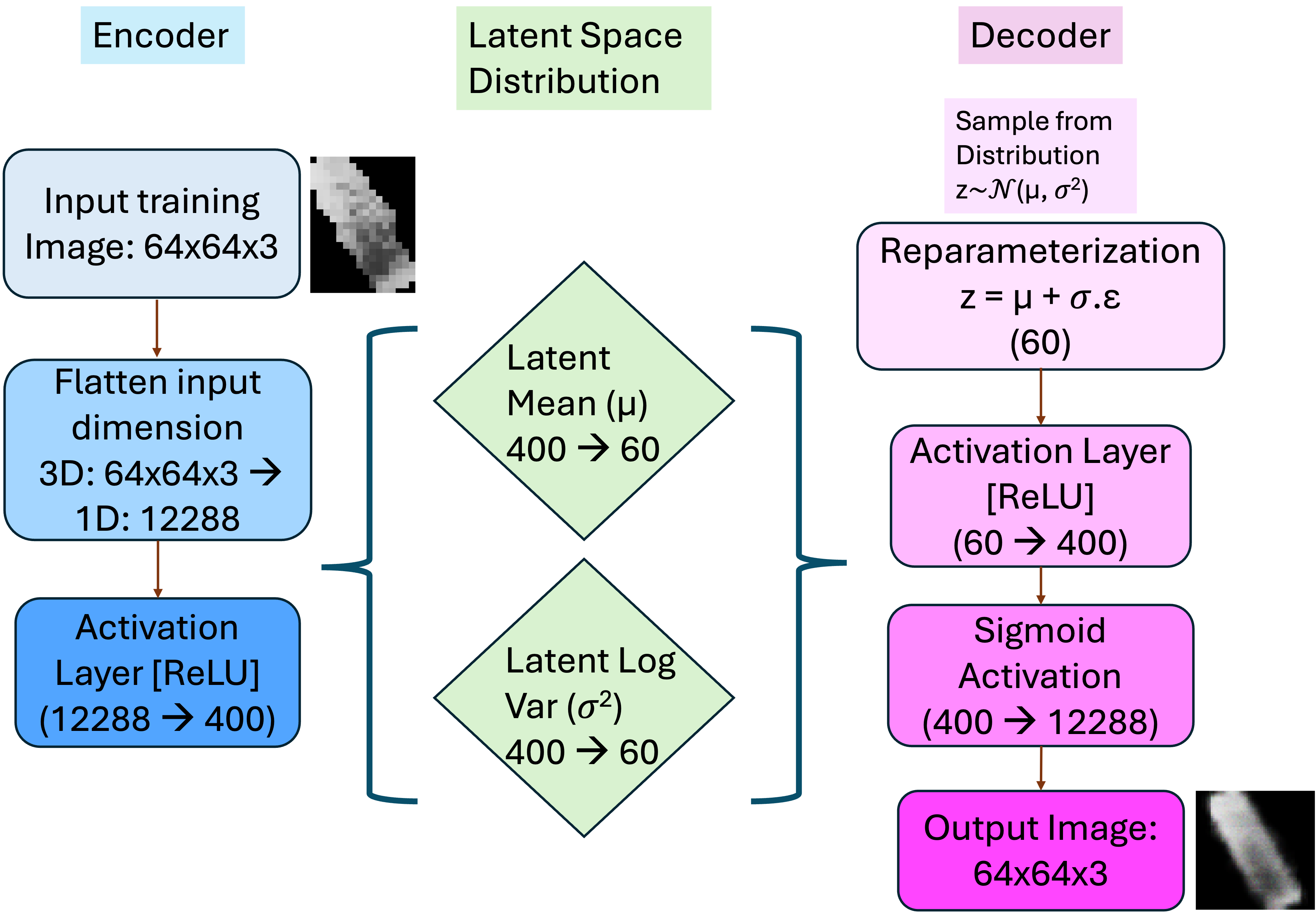}
\caption{Synthetic MFM segments are generated using VAE architecture. It consists of the following steps: The training image of dimension (H=64 x W=64 x RGB-Channel=3) is flattened into a vector of size 12288 for processing. The encoder part of the architecture consists of a fully connected (dense) layer with input size of  12288 (flattened image) and ReLU activation function outputs a compressed representation of 400 while retaining the relevant information. The encoder then splits into two separate layers to model the latent space. The mean($\mu$) of the latent Gaussian distribution represents the central tendency of the latent space for each given input.
It determines the "position" of the latent vector in the latent space of dimension 60 and the Latent log layer produces the  log variance (log$\sigma^2$) of the latent Gaussian distribution. The latent vector "z" is sampled from the Gaussian distribution defined by $\mu$ and $\sigma$ : z = $\mu + \sigma . \epsilon$,  where $\epsilon \sim N(0,1)$, is the random noise sampled from standard normal distribution. The first fully connected dense decoder layer begins to expand the compressed latent vector back into a higher-dimensional representation with ReLU activation function capturing the non-linear transformations.The Sigmoid activation of the 2nd fully connected dense layer of the decoder ensures the output is normalized matching the input image's pixel intensity range. BCE loss function measures the reconstruction loss and KL divergence regularizes the latent space by encouraging the learned distribution $q(z|x)$ to be close to the prior distribution $p(z) \sim N(0,1)$.}\label{VAE Architecture}
\end{figure}

 We implemented VAE model to learn and generate synthetic MFM images from an input dataset of segmented nanomagnets. The workflow follows the schematic architecture provided in the Fig.\ref{VAE Architecture}, representing a structured representation of encoder, latent space, and decoder operations.

The original dataset is initialized using a CSV file containing segment metadata and each image is retrieved using an index-based file naming scheme. The segment image is converted from BGR to RGB format for compatibility with the PyTorch \cite{paszke2019pytorch} module tensors. The grayscale thresholding is applied to extract relevant nanomagnetic features, normalized by scaling pixel values
[0,1] and reshaped to (C, H, W) tensor format.
The dataset is split into training (80\%) and validation (20\%) subsets to prevent the overfitting and custom collate function filters out corrupted or missing images.
The VAE follows a structured encoder-bottleneck-decoder approach:

\textbf{Encoder:} This section of the model compresses the input image into a lower-dimensional latent representation.
The input image is flattened from (64,64,3)
to a 1D vector of size 12,288.
A fully connected layer reduces the feature dimension to 400 and the activation function ReLU introduces the non-linearity. Two separate fully connected (FC) layers compute the latent mean vector ($\mu$), representing the central position of each data point in latent space, and the log variance vector ($\log\sigma^2$), indicating how much the data spreads around this position. To address the non-differentiability of directly sampling from the distribution defined by $\mu$ and $\sigma^2$, which would prevent gradient backpropagation and model training, the model employs the reparameterization trick, sampling from a standard Gaussian distribution and then scaling and shifting this sample using the learned $\mu$ and $\sigma$. The randomness is introduced only from $\epsilon$, which is independent of the model parameters. The backpropagation gradients flow through $\mu$ and $\sigma$, allowing the network to learn both the mean and variance effectively:

\begin{equation}
    z = \mu + \sigma \cdot \epsilon,
\end{equation}
where $\epsilon$ follows a standard normal distribution N(0,1). This step establishes stochasticity in image generation, allowing continuous latent space sampling.
This probabilistic space  allows the model to learn meaningful and continuous representations of the input (e.g., shape, color, or orientation of an object). Unlike standard autoencoders, which encode input into a deterministic latent vector, VAEs treat the latent space as a probability distribution which assures that the latent space is smooth and continuous, allowing for interpolation and generation of new data points. The latent dimension, chosen as 60, describes the size of each latent vector \textit{z}, the compressed representation of each sample.
The latent space features are learned representations that encode important attributes of the input data. Each dimension in \textit{z} captures different aspects of the input, such as:
global features like shape, orientation, size, or brightness of an image and local features like texture, edges, or specific patterns. In an image dataset, some latent dimensions might encode properties like image brightness in $z_1$, orientation of objects in $z_2$ or texture details in $z_3$. However, the exact meaning of each dimension is not explicitly defined but is learned implicitly during training.

\textbf{Decoder:} The latent representation, \textit{z}  (dim = 60)is passed through a fully connected ReLU layer to map it back to 400 dimensions and another fully connected layer expands it back to 12,288 dimensions. A sigmoid activation function ensures that the pixel values are bounded between 0 and 1, so that the reconstructed output can be reshaped back into the original image dimensions (64,64,3).
The loss function of the VAE model consists of two key components:\\
(a) Reconstruction Loss (Binary Cross-Entropy - BCE): The reconstruction loss measures how accurately the decoder can reconstruct the original input images from the latent representation. Specifically, the binary cross-entropy (BCE) loss quantifies the difference between the input image and the reconstructed image on a pixel-by-pixel basis. BCE loss is defined as:

\begin{equation}
\mathcal{L}_{\text{recon}} = \sum \left[ x_i \log(\hat{x}_i) + (1 - x_i) \log(1 - \hat{x}_i) \right]
\end{equation}

where, x$_i$ is the original input image's pixel value at position i, normalized between 0 and 1, and $\hat{x}_{i}$ is the reconstructed image's pixel value at the corresponding position i, also normalized between 0 and 1.
This equation computes the pixel-wise reconstruction loss, summing over all pixels in the image. Minimizing this loss term encourages the decoder to generate output images closely matching the input images.

(b) KL Divergence (KLD) Loss: The KL Divergence loss acts as a regularization term, enforcing the latent space distribution to approximate a standard Gaussian distribution ($\mathcal{N}(0,1)$). By doing so, it prevents the model from overfitting and encourages the latent space to have desirable properties, such as continuity and smoothness. The KL divergence loss is computed as:

\begin{equation}
\mathcal{L}_{\text{KL}} = -\frac{1}{2} \sum_j \left( 1 + \log \sigma_j^2 - \mu_j^2 - \sigma_j^2 \right)
\end{equation}

where  $\mu_{j}$ is the mean of the latent Gaussian distribution for the latent dimension j and $\sigma{_j}^2$ is the variance of the latent Gaussian distribution for the latent dimension j. The $\log \sigma_j^2$ is computed directly by the model, and this equation ensures that the learned latent distributions do not deviate significantly from a normal distribution, thus regularizing the learned latent representations.

The total loss function for the VAE combines both reconstruction and regularization losses as follows:

\begin{equation}
\mathcal{L}_{\text{total}} = \mathcal{L}_{\text{recon}} + \beta \cdot \mathcal{L}_{\text{KL}}
\end{equation}

where the hyperparameter 
$\beta$ controls the relative importance or balance between reconstruction accuracy and latent space regularization. Adjusting $\beta$ allows for fine-tuning of the latent representation and reconstruction quality according to specific application needs.

The VAE model is trained using PyTorch's Adam optimizer \cite{zhang2018improved}, with learning rate of 0.005, batch size of 64 for 100 epochs. 
During each epoch the images are loaded and flattened. The VAE forward pass computes latent distribution ($\mu$,$\sigma$), sampled latent vector \textit{z}, the reconstructed image.
The loss function is then calculated and backpropagated while the optimizer updates model weights to improve reconstruction quality. After training the test MFM images are passed through the VAE to generate synthetic reconstructed images. The pipeline for synthetic image reconstruction can be summarized as follows: original test images are encoded into the latent space, and then the decoder generates synthetic reconstructions.
The reconstructed images are saved and compared with the original samples. For validating  the synthetic image quality, we compare original with generated images visually as well as evaluate pixel-wise mean squared error (MSE), SSIM and absolute pixel difference  between original and synthetic samples as described in the SI(\ref{SI_Theory}), to verify that the latent space retains key magnetic features.

\subsubsection{Frustration Analysis}\label{frustration analysis method}

The generated MFM image of honeycomb ASI is then processed for classifying frustrations and also identifying nanomagnets whose directions are needed to be toggled to minimize frustrations.
The frustration classification has 3 stages:\\
(a) Identification of segment centroid \\
(b) Honeycomb vertices identification \\
(c) Classification of the vertices \\

\textbf{Identification of Segment Centroid}:\label{segment_sentroid_method}For each synthetic MFM segment, converted to grayscale, the binary threshold is applied to remove the background and isolating the nanomagnetic structure. A bounding box is computed around the segment such that the boundary matches with the original segment. The connected component mask extracts contours representing the original segment with an area above a predefined threshold of 20 pixels and the rest are discarded to avoid artifacts or partial segments. Once segment bounding boxes are established, each synthetic segment is resized to match the bounding box of its original counterpart and overlaid onto the reassembled image at the correct location as explained in the Section \ref{segmentation method}. The centroid of each segment is calculated using:

\begin{equation}
    x_{center} = x+\frac{w}{2}, y_{center} = y + \frac{h}{2}
\end{equation}

where (x,y) are the top-left coordinates of the bounding box, and (w,h) are its width and height.
Each centroid position is stored in a list alongside its segment label for reference as Segment ID, the unique identifier for each nanomagnetic segment, x-coordinate which is the horizontal position of the centroid and y-coordinate stores vertical position of the centroid. \\
\\
\textbf{Vertex Identification of Honeycomb lattice Spin Ice}: \label{vertex_centroid_method}: In an artificial spin-ice honeycomb lattice, a vertex is defined as the meeting point of three adjacent nanomagnets \cite{nisoli2013colloquium}. Each vertex corresponds to a trivalent interaction, where the three magnetic dipoles contribute to the local frustration state. So from the calculated centroid  we can deduce that each vertex must be formed by three neighboring segment centroids. To identify vertices, a maximum distance threshold (radius = 30 pixels) is applied which defines the maximum allowable distance between two segment centroids to be considered part of the same vertex.
To identify all possible vertices, our algorithm considers every unique triplet of segments from the dataset which stores the segment ID, and their corresponding centroid (x,y). Using combinatorial selection, we iterate over all segment triplets, $Triplet = \{(x_1,y_1),(x_2,y_2), (x_3,y_3)\}$, and each possible set of three segments is extracted and their centroid positions are retrieved.


For each triplet of segments, the pairwise Euclidean distances are computed to check for spatial connectivity:

\begin{equation}
    d_{12} = \sqrt{(x_2 - x_1)^2 + (y_2 - y_1)^2}
\end{equation}
\begin{equation}
    d_{13} = \sqrt{(x_3 - x_1)^2 + (y_3 - y_1)^2}
\end{equation}
\begin{equation}
    d_{23} = \sqrt{(x_3 - x_2)^2 + (y_3 - y_2)^2}
\end{equation}

If all three distances are within the predefined radius threshold (r), the segments are considered to form a valid vertex; $d_{12}, d_{13}, d_{23} \leq r$.


To calculate the vertex centroid, the vertex location is computed as the centroid of the triangle formed by the three segments:

\begin{equation}
    x_{vertex} = \frac{x_1 + x_2 + x_3}{3}, y_{vertex} = \frac{y_1 + y_2 + y_3}{3}
\end{equation}

Thus each vertex is assigned a unique central coordinate.

The energy state of each vertex is determined by its three associated segments. Therefore, we classify the vertices to identify the high-energy points within the honeycomb lattice. \\

\textbf{Frustration Classification}: In an ASI system, frustration occurs when competing magnetic interactions prevent a system from reaching a globally minimized energy state \cite{drisko2017topological}.
Each vertex in the honeycomb lattice consists of three interacting nanomagnets, forming a three-state system where the dipole orientations define its frustration state. \\

The frustration states in the VAE-generated ASI system are classified as high energy and low energy vertices. The high energy vertices are sub-classified into +3q (3-in), direction of three arrows from the associated segments convergent towards the vertex  and -3q (3-out), where the three arrows are divergent from the vertex. The low energy vertices are also sub-classified into +q (2-in, 1-out) and -q (2-out, 1 in).  Thus the analysis is based on the magnetic moments of nanomagnet segments and their interactions at lattice vertices. \\

From above sections we identified each segment and vertex positions of the VAE generated ASI system and the magnetic moment orientations are extracted from the synthetic magnetic moment dataset the method of which are explained in section \ref{magnetic moment method}. The vertex segment associations are retrieved to map each segment to its corresponding vertex and each segment's magnetic moment is then mapped to represent the magnetic moment set ($M_s$) for a given vertex consisting of three segment IDs ($S_1,S_2,S_2$): $M_s = \{ S_1, S_2, S_3\}$. For each segment, we extract its binary direction (0,1) from the dataset and assign magnetic moments pointing towards right as 0 and left as 1.\\

A vertex is classified based on the relationship between the horizontal segment ($H$), the two angled segments ($A_1$,$A_2$) and the corresponding vertex position. If  magnetic moment of the two angled segments match while the horizontal segment opposes i.e., $A_1$ = $A_2$ $\neq$ $H$, the system is unable to achieve a minimum energy state, leading to frustration and if the magnetic moment of all three segments are equal i.e, $A_1$ = $A_2$ = $H$, the system achieves a stable low-energy configuration.
No we further classify the frustrated vertex and we consider the position of vertex with respect to the horizontal segment as the reference point.
To identify the horizontal segments, we extracted the contours of the segment and apply an ellipse fit to estimate the orientation. We calculate the segment's angle ($\theta$) from its bounding contour,$S_i$ for the i-th segment, as given by:
\begin{equation}
    \theta = fitEllipse(S_i)
\end{equation}

The angle of the horizontal segment ($\theta_h$) associated with each vertex is then identified with an angle closest to 0$^\circ$ or 90$^\circ$, as

\begin{equation}
    \theta_h = min(|\theta - 0^{\circ}|, |\theta -90^{\circ}|)
\end{equation}

The frustrated vertex class might have two conditions: (a) the detected horizontal segment is shared between two frustrated vertices and (b) the frustrated vertices are disconnected. 
For the first case, if a vertex lies to the right of the horizontal segment and the magnetic moment associated with the horizontal segment is 1, the right vertex is classified as  "+3q" and assigned red squares as shown in Fig. \ref{Frustration_Analysis} and the left connected vertex is classified as "-3q", denoted by magenta squares.\\
If a vertex lies to the left to the horizontal segment and the magnetic moment of the horizontal segment ($M_H$) is 0, then the left vertex is  classified as  "+3q" and the corresponding right vertex is "-3q". This classification is based on energy asymmetry in the ASI system, where relative positions impact frustration strength.
For each vertex, the algorithm compares its x-coordinate to the x-coordinate of the horizontal segment. \\

If $x_{vertex} > x_{segment}$ and $M_{H} = 1$ : +3q and the other connected vertex is -3q\\

If $x_{vertex} < x_{segment}$ and $M_{H} = 0$ : +3q and the other connected vertex is -3q.\\

For the second case where the horizontal segment is not shared by two frustrated vertices, we remove the logic where the other connected vertex is the opposite frustration. So the logic can be simplified as: \\

If $x_{vertex} > x_{segment}$ and $M_{H} = 1$ : +3q  and \\

if $x_{vertex} < x_{segment}$ and $M_{H} = 0$ : +3q.\\

The low energy state class of vertices, where the interactions between the three segments forming a vertex are minimally strained, are also sub classified with reference to the relative position and magnetic moment of the horizontal segment. \\
If the vertex is to the left of the horizontal segment ($x_{vertex} < x_{segment}$),\\
For $M_H = 1$, the vertex is +q denoted by green star and \\
for $M_H = 0$, the vertex is -q denoted by yellow star.\\

If the vertex is to the right of the horizontal segment ($x_{vertex} > x_{segment}$),\\
For $M_H = 0$, the vertex is +q denoted by green star in Fig. \ref{Frustration_Analysis} and \\
for $M_H = 1$, the vertex is -q denoted by yellow star.\\
In some cases, low energy states may still exist even when  $A_1 \neq A_2$, meaning that magnetic moment of the two angled segments do not point in the same direction. In such cases,

If $x_{vertex} < x_{segment}$ and $M_H = 1$, it is classified as +q (Green star) and $M_H = 0$ it is classified as -q (Yellow star).\\

If$x_{vertex} > x_{segment}$ and $M_H = 0$, it is classified as +q (Green star) and $M_H = 1$ it is classified as -q (Yellow star).\\

This detailed classification method helps in distinguishing stable (low-energy) regions in the spin-ice system, which is crucial for analyzing magnetic order and energy distribution.


\section{Results}\label{sec11}

The Variational Autoencoder  model plays a crucial role in accurately identifying the magnetic moments of individual nanomagnet segments in the artificial spin ice lattice. By harnessing its ability to extract high-dimensional latent features and reconstruct the Magnetic Force Microscopy images, the VAE enables precise determination of dipole orientations. This, in turn, allows for a systematic frustration analysis of the ASI system, classifying high-energy and low-energy vertex states based on their magnetic configurations.

\begin{figure}[h!]
\centering
\includegraphics[width=1.0\textwidth]{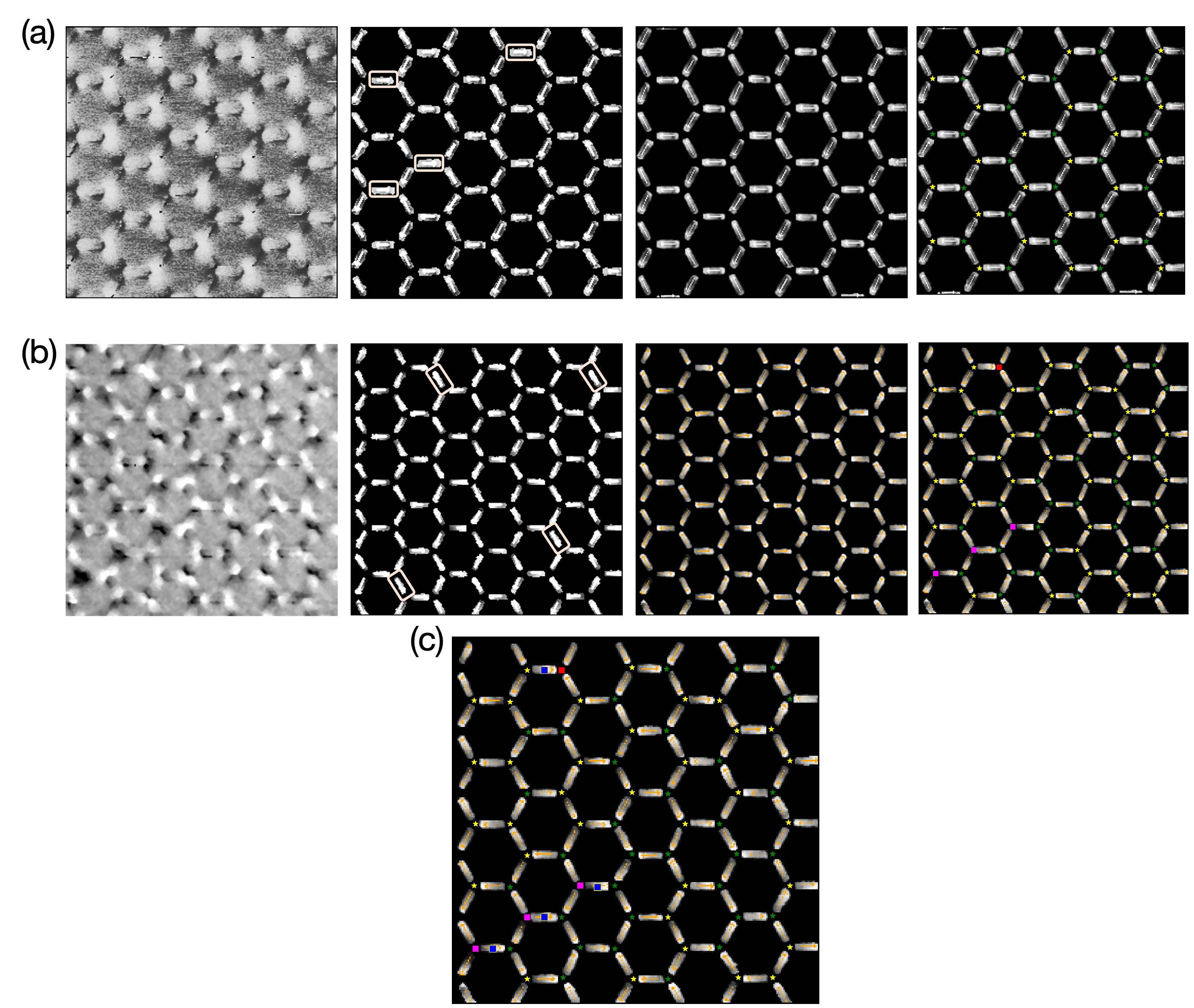}
\caption{Progression from Magnetic Force Microscopy (MFM) imaging to predicting frustration vertices in an artificial spin ice (ASI) lattice is illustrated. The process is divided into following stages:
The first row (a) presents MFM images recorded under an external magnetic field of 90 mT, applied from right to left.
The second row (b) corresponds to the relaxed state of the ASI lattice after relaxation, allowing spontaneous flipping of nanomagnet moments due to thermal activation.
The second column displays the reconstructed MFM images,pertaining to the corresponding morphology image, where individual nanomagnets are segmented and labeled, with arrows indicating the dipole moment direction.
The third column shows the synthetic nanomagnet reconstructions generated training VAE model. The dipole moment directions are calculated following the method outlined in Section \ref{method_magnetization_arrows}. The highlighted segments in the original MFM images  compared with the synthetic segments, reveals discrepancies in the top row, where miscalculations occur in moment direction, while the dipoles in the relaxed bottom row remain inconclusive.
The vertices of the honeycomb lattice are classified into high-energy monopoles and low-energy dipoles: Frustrated monopoles are further categorized into +3q (red squares) and -3q (magenta squares).
Lower-energy dipoles are classified as +q (green stars) and -q (yellow stars).
The nanomagnet segments contributing to high-energy monopoles are identified, and the frustration of the ASI lattice is iteratively evaluated. Finally, nanomagnets to be toggled for frustration minimization are detected to optimize the ASI configuration in (c).
}\label{Frustration_Analysis}
\end{figure}

The VAE model is trained on segmented nanomagnet structures extracted from MFM images, learning a compressed latent space representation of these structures. Once trained, the VAE reconstructs synthetic MFM images, preserving essential magnetic features while mitigating noise and experimental artifacts.

Figure \ref{Frustration_Analysis}(a) represents MFM images recorded under an external magnetic field of 90 mT, applied from right to left. The applied field enforces a preferential alignment of nanomagnet moments, influencing the initial spin-ice configuration. The second row \ref{Frustration_Analysis}(b) shows the relaxed ASI lattice after applying magnetic field of -25mT, where negative sign corresponds to the opposite direction of initial magnetic field condition, for relaxation period. During this time, spontaneous moment flips occur, allowing the system to transition toward a lower-energy state.
In both cases, the second column presents the segmented nanomagnets obtained from the morphology image, where each segment is labeled, and arrows indicate their magnetic dipole moment directions. \\

The third column in Fig. \ref{Frustration_Analysis} presents the VAE-generated synthetic MFM images, where individual nanomagnets are reconstructed and assigned dipole orientations based on the trained model. 
The latent space representation learned by the VAE allows for smooth reconstructions that are robust to noise and minor artifacts, serving as a method to determine magnetic moment directions with higher accuracy. \\
In Fig. \ref{Frustration_Analysis}(a), discrepancies are observed between the original and VAE-generated images. The model identifies cases where moment direction calculations were incorrect in the experimental MFM images, highlighting the advantage of using learned representations over direct image segmentation.
In the Fig. \ref{Frustration_Analysis}(b), where the ASI lattice has undergone relaxation, dipole orientations remain ambiguous in certain regions. The VAE enables a systematic approach to analyzing these fluctuations by reconstructing the most probable moment configurations given the observed data. Thus the VAE generated ASI system accomplishes two key purposes, first they were able to handle the  uncertain magnetic configurations of the original MFM images and secondly, the algorithm identifies inconclusive dipole moments of the microscopic images. As a consequential advantage, the synthetic MFM images aids in a more precise classification of lattice vertices.\\

\subsection{Optimized Spin Ice Lattice Generation}

The process of frustration minimization involves identifying and selectively toggling the magnetization direction of nanomagnets in the ASI lattice to achieve a lower-energy state while maintaining the overall structure of the spin ice. This is a two stage methodology consisting of determining magnetic moments of the segments to be toggled(flipped from 0 to 1 or vice versa) and evaluating the impact of toggling on the overall frustration state. \\

\textbf{Selecting Candidate Segments for Toggling}: The selection process prioritizes the segment that has the highest influence on multiple frustrated vertices. Each segment is classified as either horizontal or angled based on its orientation in the lattice. Horizontal segments (aligned with 0$^{\circ}$ or 90$^{\circ}$) are treated as reference points for toggling decisions. Segments associated with the highest number of frustrated vertices are given priority for toggling. If a single segment contributes to frustration in two or more vertices, toggling it may result in a greater reduction in overall frustration. The spatial arrangement of segments is another criteria of the selection process where the each segment is analyzed to ensure that toggling one segment does not introduce new frustration states elsewhere in the lattice. For each frustrated vertex, the selected segment's magnetic moment is toggled, based on the binary classification of magnetic moments as explained in section \ref{frustration analysis method} and the frustration classification of all affected vertices is re-evaluated. If toggling the segment results in a net decrease in frustration across the lattice, the change is accepted; otherwise, the segment's original state is restored.

In the toggling process, the first step is the extraction of original magnetic moment direction of the segment and temporarily toggling the segment's direction. Then we recalculate the frustration classification for all affected vertices and compare the new frustration count with the previous count. If the frustration count decreases, retain the new magnetic moment direction and if there is no improvement, revert the segment to its original direction.
The toggled segments are stored in a tracking table to ensure that future toggles do not interfere with previously optimized segments.\\

\textbf{Evaluating the Effectiveness of Toggling}: The effectiveness of the toggling strategy is measured by comparing the frustration count before and after the optimization process. The final optimized ASI configuration is saved, along with a record of which segments were toggled. The final frustration map is visualized by classifying vertices into their respective frustration categories. The segments to be toggled are marked with blue squares in Fig. \ref{Frustration_Analysis}(c) that reduces the overall frustration, generating a low energy state ASI system.
By iteratively adjusting the nanomagnet orientations, the VAE-generated ASI lattice is refined into a more stable configuration, minimizing the overall frustration in the system. The allows for precise control over frustration states in artificial spin ice systems providing a template for generating on-demand spin ice sample with the ability to reduce energy or if required, introduce high energy vertices at exact locations.





\section{Discussion}\label{Discussion}

The integration of Variational Autoencoders with Magnetic Force Microscopy imaging has enabled a novel and robust approach for analyzing frustration in artificial spin ice  systems. Traditional methodologies for determining magnetic moment orientations and frustration states often face limitations due to experimental noise, segmentation inaccuracies, and uncertainties in moment determination. The VAE-based generative model addresses these challenges by providing a data-driven framework capable of reconstructing high-fidelity synthetic MFM images, thereby allowing for a more precise identification of frustration vertices and their associated nanomagnetic configurations.

The primary advantage of employing VAEs lies in their ability to extract latent feature representations from experimental MFM data, effectively mitigating segmentation errors and compensating for inconsistencies in the direct image-based calculations. By training on segmented MFM images, the VAE learns to encode essential spatial and magnetic features, facilitating improved prediction of moment orientations even in cases where the original data is ambiguous or partially corrupted. The reconstructed synthetic images retain critical structural and magnetic information while reducing noise-induced distortions, allowing for a systematic and automated approach to frustration classification.

The capability of VAEs to generalize from learned features demonstrates that magnetic moment miscalculations present in raw experimental images are corrected, as observed in the comparison between original and synthetic reconstructions. This approach is particularly useful in identifying high-energy monopole configurations ($\pm$3q) and lower-energy dipole states ($\pm$q), enabling a more in-depth understanding of frustration dynamics within the ASI lattice.

Furthermore, the automated classification of frustration states assists in the optimization of ASI configurations through targeted toggling of nanomagnet orientations. By iteratively evaluating the impact of flipping specific nanomagnets, the framework efficiently identifies an optimal set of toggled segments that minimizes overall lattice frustration. This process provides a blueprint for designing spin-ice configurations with tailored frustration properties, offering a pathway toward controlled magnetic metamaterials with tunable energy states.

\section{Concluding Remarks}\label{Conclusion}
In this work, we have demonstrated that deep learning-based generative models, specifically VAEs, can significantly improve the accuracy and robustness of frustration classification in ASI systems. The ability to synthesize high-quality MFM images, correct for segmentation errors, and systematically classify magnetic moment orientations highlights the power of machine learning in advancing experimental condensed matter physics. The automated frustration analysis and optimization framework presented here lay the groundwork for next-generation artificial spin ice designs with precise control over magnetic configurations, contributing to the broader field of engineered magnetic materials.

\subsection{Future Direction}

The findings of this study have broad implications for the design and control of ASI-based systems in various applications, including reconfigurable magnonics, spintronics, and quantum information processing. The ability to systematically predict and minimize frustration through deep learning approaches enables the development of programmable ASI structures with precise energy landscapes.

Future work will explore expanding the VAE framework to incorporate multi-scale analysis, allowing for frustration predictions in larger and more complex ASI lattices. Additionally, integrating reinforcement learning techniques with the existing toggling strategy could further enhance the efficiency of frustration minimization by optimizing flipping sequences dynamically. Experimental validation of the proposed synthetic reconstructions through direct magnetic imaging and comparison with theoretical models will provide further insights into the reliability and accuracy of the machine learning approach.
 
\newpage

\backmatter

\section*{Supplementary Information} 
\addcontentsline{toc}{section}{Supplementary Information}
\renewcommand{\thesubsection}{S\arabic{subsection}}
\setcounter{subsection}{0}

\subsection{Fabrication of Spin Ice Sample}\label{SI Fabrication}
 The Nickel artificial spin-ice metasurface was fabricated at the Center for Integrated Nanotechnologies using a layer-by-layer approach, beginning with a silicon substrate. Initially, a 100 nm layer of gold (Au) was deposited, preceded by a 2 nm titanium (Ti) layer to enhance adhesion, utilizing electron beam evaporation for both metals. This was followed by atomic layer deposition of aluminum oxide ($Al_2O_3$) to create a thin insulating film. The top spin-ice layer of Nickel was designed using electron beam lithography, which allowed for the precise definition of a negative imprint of the final design through a dual-layer polymethyl methacrylate (PMMA) photoresist. Subsequently, an 80 nm thick layer of Nickel was deposited via electron beam evaporation, followed by a lift-off process in acetone to remove the PMMA and reveal the final structure.
\\

\subsection{Error Mapping of VAE generated and Original lattice} \label{SI_Theory}
We compared the VAE-generated ASI with magnetic moment directional arrows on each nanomagnet segment and original MFM phase image using 3 types of mappings.

\begin{figure}[H]
\centering
\includegraphics[width=1.0\textwidth]{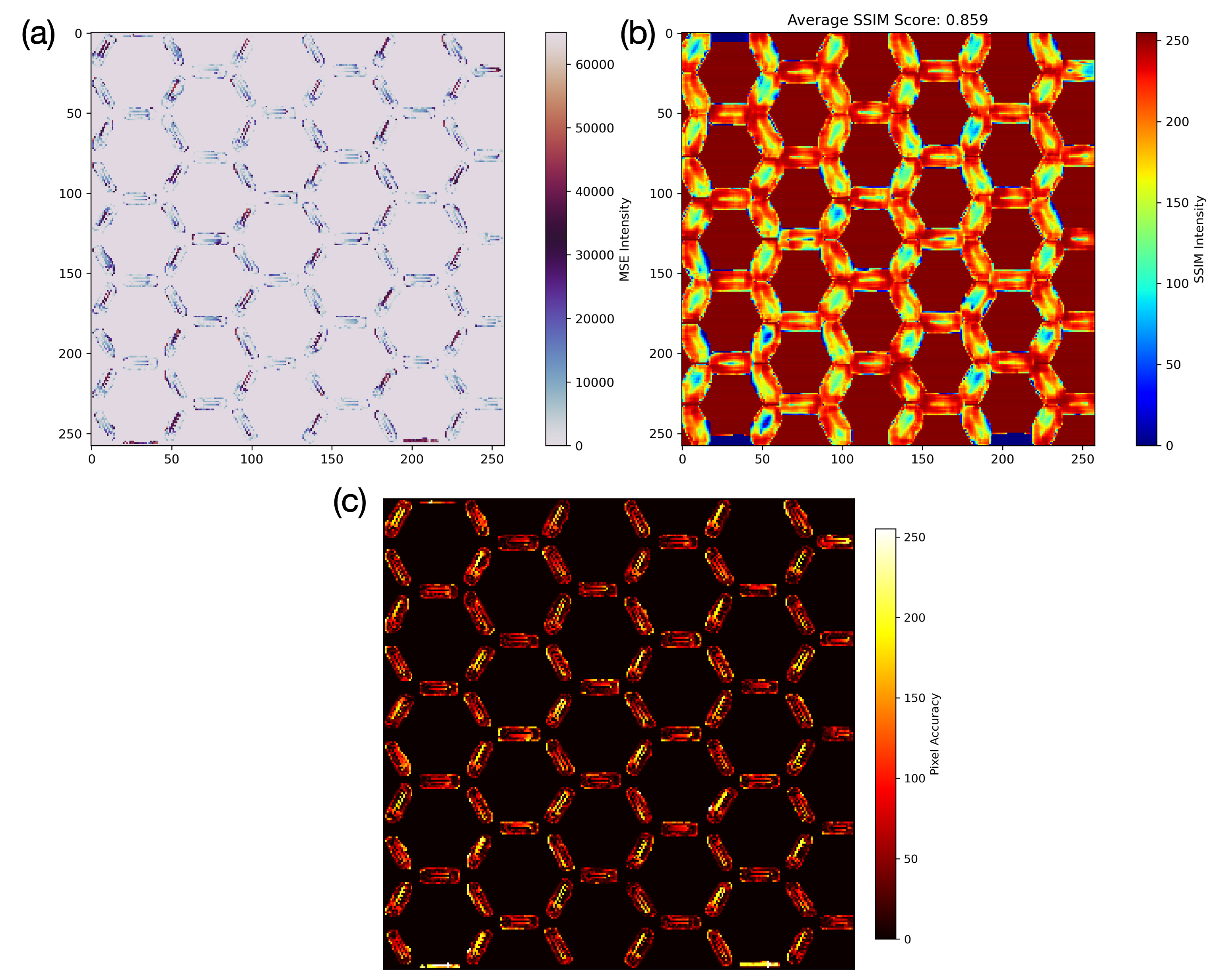}
\caption{(a) The pixel-wise Mean Squared Error between the reconstructed original ASI image and its VAE-generated synthetic counterpart is visualized in this heatmap. Each pixel's intensity reflects the squared difference between corresponding grayscale pixel values, highlighting local deviations in reconstruction fidelity.
Regions with higher MSE values (darker tones) correspond to structural or intensity mismatches, particularly along the edges and high-frequency domains of the patterned structures. Whereas the light regions indicate better agreement between the two images. The color bar on the right represents the range of MSE intensity values, scaled from 0 to above 60,000 that enables spatial localization of reconstruction errors in periodic nanomagnet arrangements, which assesses the model performance in physically structured image domains.
(b) The pixel-wise Structural Similarity Index (SSIM) map between the reconstructed original ASI image and its VAE-generated synthetic reconstruction captures local structural fidelity by evaluating luminance, contrast, and texture similarity within sliding Gaussian windows.
Each pixel's intensity in the heatmap represents its structural similarity, scaled from 0 (no similarity) to 1 (perfect match), linearly mapped to a color range for visualization (0–255). High-intensity regions (red-yellow) indicate areas of strong local structural agreement, while low-intensity regions (blue) highlight zones of poor reconstruction fidelity or feature misalignment.
The average SSIM score of 0.859,  suggests a generally strong perceptual similarity across the image, with localized discrepancies primarily along junctions and edge transitions within the patterned structure. 
(c)This heatmap visualizes the absolute pixel-wise intensity difference between the reconstructed original ASI image and its VAE-generated synthetic counterpart. Each pixel value represents the absolute grayscale intensity deviation between the two images, highlighting fine-grained local discrepancies across the spatial domain.
High-intensity regions (bright yellow) correspond to areas of greater mismatch, while darker areas (near black) indicate close pixel-level agreement. The perceptually intuitive sequential colormap allows for immediate visual localization of error-prone regions, particularly along sharp features and lattice edges where synthetic reconstructions often underperform.}
\label{Error Mapping}
\end{figure}

To identify regions with large deviations from the original mfm image, we compute average of the squared pixel intensity differences as shown in Figure \ref{Error Mapping}(a). The per-pixel squared error is calculated as follows:
\begin{equation}
    MSE_{i,j} = (I_{0}(i,j) - I_{s}(i,j))^2
\end{equation}

I$_0$ and I$_s$ represent the pixel values of original and synthetic MFM images respectively. The pixel based  MSE is raw error mapping ranging from 0-$\infty$ with lower the MSE intensity value regions more converged the VAE algorithm.

Next to correlate with human visual perception, especially for textures and patterns we computed the Structural Similarity Index (SSIM) measurement between the original and synthetic MFM images. SSIM evaluates the perceptual similarity between two images by considering changes in: Luminance (brightness), Contrast and Structure which unlike MSE, that are sensitive to pixel-level differences, is designed to align with human visual perception.
For two corresponding image patches, the original and synthetic(x,y),he SSIM is calculated as:

\begin{equation}
    SSIM(x,y)= [l(x,y)]^{\alpha} \cdot [c(x,y)]^{\beta} \cdot [s(x,y)]^{\gamma}
\end{equation}

If $\alpha = \beta = \gamma = 1$, 

\begin{equation}
    SSIM(x,y) = \frac{(2\mu_{x}\mu_{y}+ C_{1})(2\sigma_{xy}+C_{2})}{(\mu_{x}^{2}+\mu_{y}^{2}+C_{1})(\sigma_{x}^{2}+\sigma_{y}^{2}+C_{2})}
\end{equation}

The local mean($\mu_{x},\mu_{y}$), variance($\sigma_{x}^{2},\sigma_{y}^{2}$) and covariance($\sigma_{xy}$) are calculated by applying a Gassian filter which provides localized, weighted averaging for SSIM and

C$_1$ and C$_2$ are small constants to stabilize the division. The Gaussian window (or Gaussian kernel) is a 2D matrix that weights pixels according to the Gaussian distribution. It is used where local structural similarity is more important than global average. In Fig. \ref{Error Mapping}(b) we calculated the Gaussian filter window with the kernel size of 11$\times$11 for localized averaging of closer pixels rather than distant ones to measure local structural similarity. The 2D Gaussian Function is defined as 
\begin{equation}
    G(x,y) = \frac{1}{2\pi\sigma^2}exp(-\frac{x^2+y^2}{2\sigma^2})
\end{equation}

where x,y are the pixel distances from the center, $\sigma$ controls the spread and the kernel size is typically chosen as odd so that the kernel is convoluted equally on all sides from the center pixel.

And finally the one-to-one pixel comparison calculates the absolute difference between corresponding pixels in both images to produce a difference map where 0 means no difference (perfect pixel match) while higher values indicate larger pixel differences as shown in Figure \ref{Error Mapping}(c). We begin with ensuring that the images are read in grayscale mode (single-channel intensity values) to simplify pixel-wise comparisons by eliminating color channel variations. The images are resized to a smaller dimensions to avoid interpolation artifacts from upscaling.
Then after the pixel difference is calculated for both the synthetic and original MFM images, the difference map is normalized to fit within the range (0, 255) for visualization.

The Open CV attribute NORM\textunderscore{MINMAX} scales pixel values linearly between 0 (min difference) and 255 (max difference). The closer the pixel values in both images, the smaller the difference in the heatmap. Large differences indicate areas where the synthetic image fails to reconstruct the original. The pixel based error mapping helps us to assess how well the VAE has learned to reconstruct spatial patterns. The high-intensity regions in the heatmap highlight problematic areas in the VAE output.
Resizing while generating the VAE images may introduce minor distortions, but using the minimum dimensions mitigates errors. This map, unlike MSE or SSIM, provides a direct measure of pixel accuracy, making it useful for pinpointing structural shifts, boundary misalignments, or reconstruction artifacts introduced by the VAE.

\bmhead{Acknowledgements}
This work at Los Alamos was carried out under the auspices of the U.S. Department of Energy (DOE) National Nuclear Security Administration (NNSA) under Contract No. 89233218CNA000001. It was supported by Center for Integrated Nanotechnologies, a DOE BES user facility, in partnership with the LANL Institutional Computing Program for computational resources.

\section*{Declarations}


\subsection*{Conflict of interest/Competing interests}
The authors declare no competing interests.



\subsection*{Code availability } 

The code is available on request.

\subsection*{Data availability } 
Data is available on request.

\subsection*{Author contribution } 

A.N. developed all the codes for image processing, Auto-Encoder model development, frustration prediction method development and prepared the manuscript. S.M. performed the MFM measurements, generated microscopic image data, all related experiments and prepared the manuscript. P.P.I., T.M.L. and P.B. prepared the  Artificial Spin Ice sample, S.T., A.C.J. and J.X.Z. supervised the project in conception and discussion and prepared the manuscript.

\noindent

\bigskip









\bibliography{sn-bibliography}

\end{document}